\documentclass[preprint,12pt]{elsarticle}
\usepackage{epsfig}
\usepackage{latexsym,amsmath,amssymb,amstext}
\begin{document}
\title{Chemical potential on the lattice: Universal or Unique? }

\author[1]{Rajiv V. Gavai\footnote{Address after 1 January 2021 : Indian
Institute of Science Education \& Research, Bhopal bypass, Bhauri, Bhopal
462066, India.}}
\address[1]{Fakult\"at f\"ur Physik, Universit\"at Bielefeld, 
           D-33615 Bielefeld, Germany}
\ead{gavai@physik.uni-bielefeld.de}
\date{\today}
\begin{abstract}
Lattice techniques are the most reliable ones to investigate non-perturbative
aspects of quantum chromodynamics (QCD) such as its phase diagram in the
temperature-baryon density plane.  They are, however, well-known to be beset
with a variety of problems as one increases the density.  We address here the
old question of placing the baryonic (quark) chemical potential on the lattice.
We point out that it may have important consequences for the current and future
experimental searches of QCD critical point.

\end{abstract}
\maketitle
\newpage
\section{Introduction}
The behaviour of strongly interacting matter, described by Quantum
Chromodynamics(QCD), at nonzero temperatures or baryon densities has continued
attracting attention both theoretically and experimentally for more than three
decades ~\cite{pr86,qm19,xq19}.  Since QCD coupling is known to be large at or
near the scale of QCD, $\Lambda_{QCD}$, investigating the QCD phase diagram
necessitates strong coupling techniques.  Lattice QCD is the most successful
non-perturbative technique which has provided us with key interesting results
pertaining to the phase diagram. For instance, it is known from independent
lattice studies that the transition from the hadron phase to the quark gluon
plasma phase at zero baryon density is a crossover ~\cite{tc1,tc2,tc3}.
Extending these results to non-zero baryon density, or equivalently nonzero
quark chemical potential $\mu$, one encounters the famous sign problem : the
quark determinant becomes a complex number, inhibiting the use of the trusted
importance sampling based Monte Carlo methods.

Several ways have been proposed to confront the sign problem in QCD
\cite{fodor1,imagmu, cano1,taylor}.  Based on an analysis of model quantum field
theories with the same symmetries as two light flavour QCD ~\cite{pw,bpv}, a
critical end-point is expected to exist in the QCD phase diagram.  One expects
the baryon number susceptibility to diverge~\cite{gg1} there.  Consequently, its
Taylor series expansion at finite baryon density would have a finite radius of
convergence, leading to an estimate of the location of the critical
end-point~\cite{gg1,gg2}.  First such estimates of the radius of convergence of
the Taylor series suggested the critical end-point to be  at $T_E/T_c=0.94$ and
$\mu_B/T_E=1.8(1)$ ~\cite{gg2}.  A study on a finer lattice refined the
continuum limit to be around $T_E/T_c=0.94(1)~,~\mu_B/T_E=1.68(5)$~\cite{dgg}.
On the other hand, other approaches, such as employing imaginary chemical
potential and/or 'improved' actions, have reported only bounds on the location
of critical point which at 1-$\sigma$ level disagree~\cite{bibnd} with the
results of ~\cite{dgg}. In heavy-ion experiments at RHIC, the fluctuations of
the net proton number are employed as a proxy for the net baryon number. The
STAR experiment at Brookhaven National Laboratory has measured the fluctuations
of the net proton number up to the fourth order for a wide range of center of
mass energy $\sqrt s$.  At $\sqrt s=19.6$ GeV the experimental data are seen
\cite{star,star1} to deviate maximally from the predictions of the proton
fluctuations for models which do not have a critical end-point, and are similar
to the lattice QCD-based predictions \cite{ggplb} for a critical point.  While
these above mentioned results employ kurtosis of baryon/proton number, it has
been proposed that the 6$^{th}$-order fluctuations may shed light on whether the
crossover at zero baryon density is a shadow of the $O(4)$ criticality in the
chiral limit~\cite{kared}.  Clearly, still higher orders will eventually need to
be computed for better control over the radius of convergence.  Thus higher
order susceptibilities are, and will continue to remain, of immense interest.

In this paper, we compare and contrast the different ways of introducing the
chemical potential on the lattice, and assess their impact on these higher order
susceptibilities which also govern the coefficients of the Taylor series.
Astonishingly, we find that the results depend on the way chemical potential is
introduced.  The differences appear to persist in the continuum limit.  This
observation also has consequences for all other methods to tackle the sign
problem.  We argue for a choice closest to the continuum QCD as the best.  In
section \ref{sec:univ},  we recall the existing methods to place chemical
potential on the lattice and demonstrate their failure with universality. The
next section \ref{sec:conq} is devoted to a discussion of their other
attributes. We finally summarise our results.

\section{Universality and Chemical Potential}
\label{sec:univ}
\bigskip

The lattice QCD partition function in the path integral formalism is given by
\begin{equation}
 \mathcal{Z}=\int \mathcal{D}U_\mu \mathcal{D}\bar \psi \mathcal{D} \psi
\rm{e}^{-S_G -S_F(ma,\mu a)}~,~
\end{equation}
where $\psi(x)$, $\bar \psi(x)$ and $U_\mu(x)$ represent the quark, anti-quark
at site $x$ and the gluon field on the link $(x,\hat \mu)$ respectively.  $S_G$
denotes a suitable choice for the gluonic action and $S_F$ is the quark action.
We shall consider below the na{\i}ve quark action but our considerations are
easily generalized to other local actions such as the Wilson action, the
staggered action or their improved versions. Similarly, we will consider only a
single flavour with the baryonic chemical potential $\mu_B = 3 \mu$ for
simplicity, generalization to more flavours again being straightforward.
Denoting by $ma$ the quark mass and by $\mu a$ its chemical potential, the
fermionic action is given by $S_F = \bar \psi M(ma,\mu a) \psi$ with M defined
as below:
\begin{eqnarray}
\label{qm}
 S_F (ma, \mu a) &=& \sum_{x,\mu=1}^3  \bar \psi(x) \gamma_\mu \big[ U_\mu(x)
\psi(x+\hat \mu) - U_\mu^\dagger(x-\hat \mu) \psi(x -\hat \mu)\big]  \\ \nonumber
&+&\sum_{x}  \bar \psi(x) \gamma_ 4 \big[ f(\mu a) \cdot U_4(x)
\psi(x+\hat 4) -  g(\mu a) \cdot U_4^\dagger(x-\hat 4) \psi(x -\hat 4)\big]
\\ \nonumber
 &+& ma~\bar \psi(x) \psi(x)~,
\end{eqnarray}
Three possible choices have so far been used in the literature \cite{hk,kogut,
bilgav} for the functions $f$ and $g$, denoted below by subscripts 
$L$(linear), $E$(exponential) and $S$(square root):
\begin{eqnarray}
\label{prsc}
f_L (\mu a) &=& 1 + \mu a~,  \qquad\qquad\qquad\qquad g_L(\mu a) = 1 -\mu a\\ \nonumber
f_E (\mu a) &=& \exp(\mu a)~, \qquad\qquad\qquad\qquad g_E(\mu a) = \exp (-\mu a)\\ \nonumber
f_S (\mu a) &=& (1 + \mu a)/\sqrt{1-\mu^2 a^2}~, \qquad g_S(\mu a) = (1 -\mu
a)/\sqrt{1-\mu^2 a^2}~.~
\end{eqnarray}
Following the natural route of obtaining the conserved charge from the
corresponding current conservation equation on the lattice leads to the na{\i}ve
linear choice \cite{bilgav} above.  However, it has $\mu$-dependent quadratic
divergences in the number density and the energy density even for the free quark
gas. These can be eliminated by the other two options for $f$ and $g$.  Indeed
all functions satisfying $f(\mu a) \cdot g(\mu a)=1$ eliminate \cite{gavai}
those divergences.  It is a straightforward exercise to check that all these
actions lead to the {\em same} continuum action in the limit of vanishing
lattice spacing $a \to 0$ since their contribution to the action is formally of
higher order in $a$.  In principle, this should ensure that the pressure is
identical for all of them, {\em provided the $a \to 0$ limit is taken}.  The
catch though is that mostly various derivatives of the pressure are evaluated
and their computations are also necessarily performed for $a \ne 0$.  This turns
out to have serious consequences as the $k^{th}$ derivative of pressure can, and
does, carry the memory of the terms up to ${\cal O}( \mu^k a^k)$ even as $a \to
0$, as we show below.

Integrating the Grassmannian quark and antiquark fields,
one has 
\begin{equation}
 \mathcal{Z}=\int \mathcal{D}U_\mu \rm{e}^{-S_G }~{\rm Det} M(ma, \mu a) ~.~
\end{equation}

A derivative of $\mathcal{\ln~Z}$ with $\mu$ leads to the
quark number density, or equivalently (1/3) the baryon number density, defined
by,
\begin{eqnarray}
 n &=& \frac{T}{V}\frac{\partial \ln \mathcal{Z}}{\partial
\mu}|_{T=\text{fixed}} \\ \nonumber
  &=& \frac{1}{N_t N_s^3 a^3} \langle {\rm Tr} M^{-1} \cdot M' \rangle ~,~
\end{eqnarray}
where $M'$ is the derivative of the fermionic matrix $M$ with respect to $\mu
a$, T = $(N_t a)^{-1}$ is the temperature and $V = N_s^3 a^3$ is the volume.  In
the process of obtaining predictions for the signals of either the critical end
point or the two-flavour chiral transition, one evaluates higher order
derivatives of $n$ to obtain various fluctuations such as the variance, skewness
or kurtosis etc.  In fact, coefficients of $\mu^8 a^8$ have been computed in
attempts to locate the QCD critical point ~\cite{gg2}, and those of $\mu^6 a^6$
terms are expected to assist~\cite{kared} in pinning down the hints of a
critical point in the chiral limit of the two-flavour theory in the heavy ion
collision data.

In general, a ${\cal O}( \mu^k a^k)$ will clearly involve up to $k$-th
derivative of the fermion matrix $M$, and thus of $f$ and $g$.  Using the
condition $f(\mu a) \cdot g(\mu a)=1$ along with the obvious $f(0)=g(0)=1$ and
$f'(0)=-g'(0)=1$ (to ensure the $\mu N$ form in the $a \to 0$ limit) conditions,
one finds $f''(0)+g''(0)=2$.  Using the fact that particle-antiparticle symmetry
implies $f(\mu a)=g(- \mu a)$, one finds that the $f^k(0)= (-1)^k g^k(0)$, and
thus  $f''(0)=g''(0)=1$. Both $f_E$ and $f_S$ satisfy this.  Unfortunately they
differ in all the higher derivatives.  There are no more conditions to fix the
higher derivatives. Indeed, $f''''(0) = 4f'''(0) -3$ is the only new relation
one has. It is easy to verify from eq. (\ref{prsc}) that $f'''_E(0)=1$ with
$f_E''''(0) = 1$ and $f'''_S(0) = 3$ with $f_S''''(0) = 9$ do satisfy this
relation. Thus only the first derivative is identical for all the $f$'s in
eq.(\ref{prsc}). Already the second derivative $f_L''(0) = 0$ but the second
derivative is identical for $f_E$ and $f_S$ and is unity.  All further higher
derivatives are different. Note these are all pure numbers, {\em i.~e.}, an
approach to continuum limit will not change these derivatives themselves.  This
has consequences for the various higher order fluctuations of the conserved
charge.  They too will be different depending upon the choice of $f$ from
eq.(\ref{prsc}) with no hope of their converging in the continuum limit.  {\em
A~priori} all $f$ are on the same footing. This therefore appears to be then a
serious violation of universality, as $f'''$ and $f''''$ enter experimentally
measurably quantities such as kurtosis or the $\chi^B_6$.

One ought to have seen this coming after all since it is well-known
\cite{bilgav} that $f_L$ has quadratic $\mu$-dependent divergences but the 
other two do not. An easy way to see this is to
look at the expression for quark number susceptibility. It is given by
\begin{equation}
   \chi =  \frac{1}{N_t N_s^3 a^2} \left[\left\langle\left({\rm Tr} M^{-1} M'\right)^2
                     \right\rangle +
         \left\langle{\rm Tr}
             \left(M^{-1} M'' - M^{-1}M'M^{-1}M'\right)
                  \right\rangle\right].
\label{chi}
\end{equation}

Since $f''_L = g''_L = 0$ for the na{\i}ve linear choice for all $\mu a$, the
first term in the second expectation value vanishes whereas $f''(0) = g''(0) =
1$ leads to a nonzero contribution from it for the other two actions. Indeed, it
is precisely this term which ensures elimination of the divergence for them.  It
is important to note that the second derivative comes from ${\cal O}(\mu^2 a^2)$
terms in $f_E$, $f_S$ and $g_E$, $g_S$ in eq.(\ref{prsc}). These by themselves
are irrelevant terms for the continuum limit of the pressure itself.   However,
they are not irrelevant at the susceptibility level. In particular, all of the
terms enclosed in the square brackets in eq.(\ref{chi}) are parametrically
independent of $a$ and contribute formally equally.  Each must vanish as $a^2$
in the continuum limit in order that $\chi$ is nontrivial in that limit.
Nevertheless, if one were to assume hypothetically that the $\langle M^{-1} M''
\rangle $ term, which is the only term to have contribution from the $\mu^2 a^2$
term, itself somehow vanishes faster than the other terms in the $a \to 0$
limit, it is clear then that the result will be the same as the linear case for
which $\chi$ is divergent.  We know, on the other hand, that the divergence in
the $a \to 0$ limit is eliminated for the exponential and square root actions
precisely due to the continued contribution of this term in the continuum limit,
as demonstrated in \cite{bilgav} for the free quark gas and in \cite{ggq} for
the interacting case.  

It should not come as a surprise that this
phenomena recurs for higher order susceptibilities as well.  One encounters even
more prescription dependence at higher orders. Consider for example the fourth
order susceptibility~\cite{gg1} : 
\begin{equation}
\label{chi4}
\chi^{4}= \frac{1}{N_t N_s^3} \left[ \biggr\langle \mathcal{O}_{1111}+6 \mathcal{O}_{112}+4 \mathcal{O}_{13}+3 \mathcal{O}_{22}+\mathcal{O}_4\biggr\rangle
- 3 \biggr\langle \mathcal{O}_{11}+\mathcal{O}_2\biggr\rangle^2\right]~. 
\end{equation} 
Here the notation $\mathcal{O}_{ij\cdots l}$ stands for the product, 
$\mathcal{O}_i\mathcal{O}_j\cdots O_l$. The relevant $\mathcal{O}_i$ for
eq.(\ref{chi4}) are~\cite{gg1}
\begin{eqnarray}
\label{Ogg}
\mathcal{O}_1 &=& \text{Tr ~} M^{-1}M', \\ \nonumber
\mathcal{O}_2 &=& -\text{Tr ~}  M^{-1}M'M^{-1}M' + \text{Tr ~} M^{-1}M'', \\ \nonumber
\mathcal{O}_3 &=& 2 ~\text{Tr ~} (M^{-1}M')^3  - 3 ~\text{Tr ~} M^{-1}M'M^{-1}M'' + \text{Tr ~} M^{-1}M''', \\ \nonumber
\mathcal{O}_4 &=& - 6~  \text{Tr ~}(M^{-1}M')^4 + 12 ~\text{Tr ~} (M^{-1}M')^2 M^{-1}M'' -3 ~\text{Tr ~}
(M^{-1}M'')^2 \\ \nonumber
&-& 3 ~\text{Tr ~} M^{-1}M'M^{-1}M''' + \text{Tr ~} M^{-1}M''''.
\end{eqnarray}

Note that there is no explicit $a$-dependence on either side of the equation
(\ref{chi4}).  $\chi^4$ is a pure number in the continuum limit. The traces run over
all the lattice sites, get normalized by the lattice volume, and are thus pure
numbers as well. Again {\em a priori} none of the trace terms vanishes faster
than any other in the continuum limit.  Since $\mathcal{O}_3$ and
$\mathcal{O}_4$ have terms with $M'''$ and $M''''$, which in turn contain the
$f'''$, $g'''$, $f''''$ and $g''''$, it is clear that for each choice out of the
three in eq.(\ref{prsc}), one will obtain a {\em different~value} on the {\em
same} set of dynamical gauge configurations.  It is worth reminding that the
differences due to the actions are pure numbers. For instance, $f_E'''(0) = 1$,
$f_E''''(0) = 1$ while $f'''_S(0) = 3$ with $f_S''''(0) = 9$.  Thus $\chi^4$
onwards for all higher order susceptibilities one obtains results which depend
on the choice of $f$ and $g$ and are thus not universal.   Since $\chi^n \propto
a^{(n-4)}$, the overall behaviour of the corresponding RHS will naturally scale
appropriately with $a$, as seen for $\chi$ in eq.(\ref{chi}).  But all the
individual terms of RHS can, and do, scale that way with none destined to vanish
faster than the others as all are formally $a$-independent.  Therefore, the
three actions continue to give different results for all $n \ge 5$.  It may
happen that the corresponding $\chi^n$, $n \ge 5$, vanishes in the continuum
limit, as for the ideal quark gas, yielding then action independent results.  In
the interacting case though this is not the case in general.

This loss of universality is not limited only to the higher order fluctuations
of the conserved charges computed using lattice QCD simulations.  Recall that
the pressure $P$ is usually constructed as a series in $\mu^B $ with these
susceptibilities as the coefficients. Hence, the pressure, and consequently all
thermodynamic quantities derived from it, are also similarly prescription
dependent from the fourth order onwards. If one somehow computed the partition
function itself directly, and obtained the continuum limit of the 
correspondingly evaluated pressure as its logarithm then it would not 
suffer from this loss of universality. On the other hand, any method which 
relies on computations of thermal expectation values of quark number 
susceptibilities will suffer from this problem.

In short, the quest to get rid of the $\mu$-dependent divergences lead to
modification of the action in the Euclidean representation of the partition
function, ostensibly by adding terms which are {\em irrelevant} in the continuum
limit $a \to 0$.  The presence of the dimensional parameter $\mu$ in these
terms, however, spoils this na{\i}ve expectation of universality for quark
number susceptibilities. Employing the
$f_L$ and $g_L$ prescription has the advantage of being faithful to the
continuum theory in reproducing the higher order fluctuations, but also has the
disadvantage of a $\mu$-dependent divergence, again as in the continuum theory.

\section{Conservation of Charge}
\label{sec:conq}
\bigskip

Recall that invariance of an action under a global $U(1)$ symmetry leads to a
current conservation equation, $\partial_\mu j^\mu(x) = 0$, and hence the
conserved charge $N = \sum_{\vec x} j^4( \vec x)$. It is worth noting that one
can compute the current conservation equation after the addition of $\mu \bar
\psi(x) \gamma_4 \psi(x)$ term in the Lagrangian to find that the additional
term  does not alter the above current conservation equation, with the conserved
charge remaining the same as it should.

For the lattice theory one can similarly demand invariance of eq.(\ref{qm})
under the global $U(1)$ symmetry: For $\psi' = \psi + \delta \psi$ and $ \bar
\psi' = \bar \psi + \delta \bar \psi$, $\delta S_F = 0$, where  $\delta \psi  =
i \epsilon \psi$, and $\delta \bar \psi  = - i \epsilon \bar \psi$ and
$\epsilon$ is small.  The resultant current conservation equation is easily
worked out as $\sum_{\mu} [j^\mu(x - \hat \mu) - j^\mu(x)] = 0$ for the case
$\mu a=0$ when $f$ and $g$ are unity in general.  Here $j^\mu(x) = \big[ \bar
\psi(x) \gamma_\mu U_\mu(x) \psi(x+\hat \mu) + \bar \psi(x+\hat \mu) \gamma_\mu
U^\dagger_\mu(x) \psi(x)\big] $ is the point split version of the usual current
one obtains in the continuum theory.  The conserved charge is related to this
point split $j^4$ in the same way and the addition of $\mu N$ term leads to the
linear action with $f_L$ and $g_L$.  For the case of $\mu \ne 0$, one can again
follow the same procedure to work out the new current conservation equation.

A simple trick to write the generic $f$ and $g$ as $\big[(f+g)/2 \pm (f-g)/2 
\big]$ respectively makes it easy to follow the derivation.  The
$\delta S_F = 0$ equation can then be simplified similarly with two differences.
$\delta S_F$ has an additional term proportional to $[f(\mu a) - g (\mu a)]/2$,
which is given by
\begin{eqnarray}
\label{dsadd}
 \delta S_F^{add} (ma, \mu a) &=& [f(\mu a) - g (\mu a)]/2 \sum_{x}  \big[ \bar \psi(x) \gamma_4 U_4(x)
\psi(x+\hat 4) \\ \nonumber &+& \bar \psi (x) \gamma_4 U_\mu^\dagger(x-\hat 4) \psi(x -\hat 4)  \\ \nonumber
&-&  \bar \psi(x - \hat 4) \gamma_ 4 U_4(x- \hat 4) \psi(x) -  \bar \psi (x +
\hat 4) \gamma_4 U_4^\dagger(x) \psi(x)\big]~.
\end{eqnarray}
Noting that $x$ is a dummy sum variable, and substituting $y= x \pm \hat 4$ in
the two terms on the third line of the eq. (\ref{dsadd}), it is easy to show
that $ \delta S_F^{add} (ma, \mu a)= 0$. Secondly, the current conservation form
of the full $\delta S_F=0 $ has a difference with the $\mu=0$ case.  The
expression for $j^4(x)$ is replaced by $j_{mod}^4(x) = ~[f(\mu a)+g(\mu
a)]/2~\big[ \bar \psi(x) \gamma_4 U_4(x) \psi(x+\hat 4) + \bar \psi(x+\hat 4)
\gamma_4 U^\dagger_4(x) \psi(x)\big] $~, resulting in the modified conserved
charge being $N_{mod} = \sum_{\vec x} j_{mod}^4(\vec x)$.  Substituting the $f$
and $g$ from eq. (\ref{prsc}), one can work out the consequences in each case.
For the linear case, $N_{mod}= N$ and thus remains unchanged. For the other two
cases, namely the exponential and the square root forms, $N_{mod}$ itself is
$\mu$-dependent for nonzero $a$, being $\cosh (\mu a)N(\mu a =0)$ and $N (\mu a
=0)/\sqrt{1 -\mu^2 a^2}$ respectively.  While in the continuum limit these
functions can be expanded to obtain a quadratic $a$-approach to the standard
conserved charge, the modifications persist for any finite $a$ employed in the
usual simulations.   A conserved charge being dependent on the chemical
potential itself may be dismissed as a lattice artifact but
this does becomes a problem in defining the canonical partition
function on the lattice.

Using fugacity $z = \exp(\mu/T)$, one relates the grand canonical partition
function to the canonical ones : ${\cal Z^{GC}} = \sum_n z^n {\cal Z}^C_n $.
Since  $z_{lat} = \exp{(N_t \mu a)} $, such a relation is feasible only for the
linear prescription of adding chemical potential. As one notices the conserved
number ought to remain the same $\mu$-independent constant for such a relation.
Since $N_{mod}$ depends on $\mu$ for the exponential case, a canonical ensemble
with constant $N_{mod}$ is not easily defined from the ${\cal Z^{GC}}$.  Indeed,
${\cal Z^{GC}} \ne \sum_n z_{lat}^n {\cal Z}^C_n $ in that case. This is of
course also the case for all $f$ and $g$ which satisfy $f \cdot g = 1$

\section{Summary}
\bigskip

Current and future experimental programs on heavy ion collisions aim to measure
fluctuations of conserved charges precisely.  The STAR results already exhibit
intriguing structure in higher order proton number fluctuations such as
kurtosis.  Still higher order fluctuations ($\chi^B_6$) are anticipated to shed
light on the nature of the chiral phase transition. Reliable theoretical
predictions are needed for these for a trustworthy comparison. Lattice QCD at
finite density is the best tool one currently has.

Defining a conserved charge, for instance the baryon number, from the
corresponding conserved current defined on the lattice and adding it using the
canonical Lagrange multiplier type linear chemical potential term in the fermion
actions on the lattice is most natural. Its $\mu$-dependent divergences lead in
the past to the proposals of other action, including the popular exponential
action.  We showed that these actions lead to {\em different} results for the
{\em same} physical quantities, namely the higher order fluctuations starting
from skewness, kurtosis and so on. These differences in the same physical
quantity persist in the continuum limit of $a \to 0$, and therefore the actions
designed to eliminate free theory $\mu$-dependent divergences violate
universality. Only the action linear in $\mu$ has an unchanged current
conservation equation and hence the same conserved charge for $\mu \ne 0$, as is
also the case in the continuum theory. Other actions, including the popular
exponential form, do not share these properties: the conserved charge on the
lattice itself becomes a function of $\mu$. The usual definition of canonical
partition function therefore does not hold for these forms. It would be an
interesting challenge to  devise  a suitable compact form to define it
on the lattice.

  It may be worth noting that preservation of exact chiral invariance
on the lattice seems feasible only for a linear form ~\cite{sharma} for the
continuum-like overlap and the domain wall fermions. Since a $\mu$-dependent
divergence exists already in the continuum for a gas of free fermions, and is
subtracted there, it can similarly be subtracted out in simulations~\cite{GaSh}.
Action with linear chemical potential term is thus unique in that it mimics the
continuum behaviour faithfully for both local and nonlocal fermion actions.
Modifying the local action to eliminate the divergence leads to a loss of
universality for higher order susceptibilities.

\section{Acknowledgements} \bigskip

It is a pleasure to gratefully acknowledge the support by the Deutsche
Forschungsgemeinschaft (DFG, German Research Foundation) through the the CRC-TR
211 'Strong-interaction matter under extreme conditions'– project number
315477589 – TRR 211. The author is also very happy to acknowledge the kind
hospitality of the Physics Department of the Universit\"at Bielefeld, Germany, 
in particular that of Profs. Frithjof Karsch and Olaf Kaczmarek.

\end{document}